\documentclass[epj,referee]{svjour}
\input epsf
\input rotate
\usepackage{graphicx}

  \newcommand{\drp}[2]{\frac{\partial #1}{\partial #2}}

\begin{document}
\title{Dissipative flows of 2D foams}

\author{Cantat I.\inst{1} \and  Delannay R.\inst{1} 
}                                                      

\institute{ \inst{1} GMCM, Universit\'e de Rennes (CNRS),
             Campus de Beaulieu  B\^at. 11A
                         CS 74205
                  263, av. du G\'en\'eral Leclerc
                     35042 Rennes Cedex, France
		    }
\date{Received: date / Revised version: date}
%
\abstract{
We analyze the flow of a liquid foam between two plates separated by 
a gap of the order of the bubble size (2D foam). We concentrate on 
the salient features of the flow that are induced by the presence, in 
an otherwise monodisperse foam, of a single large bubble whose size 
is one order of magnitude larger than the average size.
We describe a model suited for numerical simulations of flows of 2D 
foams made up of a large number of bubbles. The numerical results are 
successfully compared to analytical predictions based on scaling 
arguments and on continuum medium approximations.
When the foam is pushed inside the cell at a controlled rate, 
two basically different regimes occur: a plug flow is observed 
at low flux whereas, above a threshold, the large bubble migrates 
faster than the mean flow.
The detailed characterization of the relative velocity of the large 
bubble is the essential aim of the present paper. The relative 
velocity values, predicted both from numerical and from analytical 
calculations that are discussed here in great detail, are found to be 
in fair agreement
with experimental results from \cite{cantat05b}.\\
\PACS{82.70.Rr, 83.50.Ha, 83.60.La
      } 
} 
\maketitle
\section{Introduction}
Among complex fluids and structured materials, fluid foams, which are 
primarily materials of outstanding industrial importance, deserve too 
a particular attention as
model systems.
Indeed, they are governed by relatively well understood and simple
local equilibrium properties. Further, each individual cell of a 2D foam
  can be followed during its motion and its deformation. The relation 
between the local structure
and the induced macroscopic visco-elastic flow can thus be directly observed.
These reasons explain why
theoretical studies
of 2D foam flows in viscous regimes are of growing importance
\cite{schwartz87,reinelt89,glazier92,li95,okuzono95,durian95,cantat03a,kern04}.

\begin{figure}[h]
\centering
\includegraphics[width=6cm]{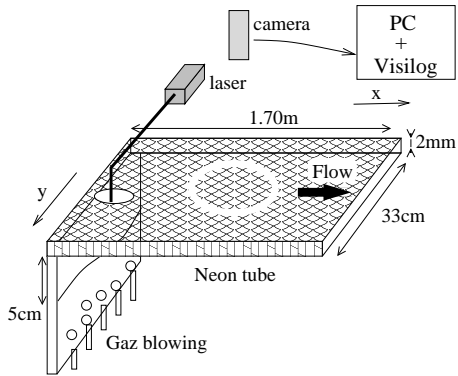}
\caption{\it  Schematic drawing of the set-up from which we collected 
the experimental results \cite{cantat05b} referred to in the present 
theoretical paper.
}
\label{experience}
\end{figure}

By a 2D
foam, we refer here to foams confined between two horizontal plates and made up
of a single bubble layer.
The richness of the flow properties is largely related
  to the disorder of the foam structure \cite{sollich97}, but we shall show in
this paper that even few defects in an otherwise basically ordered
structure may
already induce strong
  modifications of the flow in a viscous regime.

  If a constant foam flux
is imposed in
  a Hele-Shaw cell (see Fig.\ref{experience}), a uniform pressure 
gradient is created between upstream and
downstream
  to compensate the viscous drag on the plates.
Lateral boundary effects are negligible and the reference flow at low
velocity is thus a
simple plug flow.
At higher velocity, the plug flow becomes unstable for polydisperse foams.
The velocity field, then dominated by the bubble size distribution and by its
spatial correlations, leads to a
very complex and fluctuating flow. Larger bubbles move faster than
the smaller  ones and the
foam structure is fully reorganized with strong attractive 
correlations  between
bubbles of similar sizes.
An understanding of this surprising foam behavior is out of reach
without a satisfactory
description of the underlying
elementary instability which involves a single defect. Preliminary 
results on this instability were
shortly discussed in a previous article \cite{cantat03a}, and 
experimental results are thoroughly described in  \cite{cantat05b}.
  In the following, we shall report in detail on theoretical and numerical
investigations of the flow of a 2D monodisperse foam in which a single
large bubble is embedded, leading to what is called hereafter the 
{\it large bubble instability}.

\begin{figure}[h]
\centering
\includegraphics[width=5cm,angle=-90]{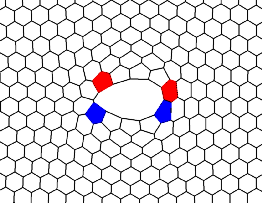}
\includegraphics[width=5cm,angle=-90]{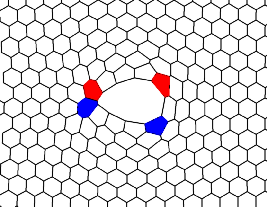}
\caption{\it Two 'shots' of the LB migration through the foam, in 
the LB frame (numerical simulations). The foam crystalline orientation is the most favorable and the
migration thus occurs in the same direction as the mean flow, to the right.  
}
\label{migra}
\end{figure}

The viscous dissipation is localized on the Plateau borders touching
the plates,
{\it ie} the
contact lines between the vertical foam films and the films wetting the plates.
In a monodisperse foam which flows uniformly, viscous forces, 
averaged at the scale of few
small bubbles, induce a uniform
stress field oriented upstream. For a large bubble (denoted by LB in 
the following)
the density of Plateau borders is smaller, and the averaged viscous 
stress is lower.
  The LB is, in a way, analogous to a
low density drop embedded in a fluid submitted to a uniform gravity field.
In both cases, the uniform stress field is balanced by a linear 
pressure field, and a resulting
   Archimedes-like
force is exerted on the bubble (or respectively on the drop)  in the  direction 
opposite to that of the force field.
  This driving force, responsible for the LB migration,
competes with the
elastic reaction of the bubble network : a stable foam structure is 
obtained at low
velocity. By contrast, the large bubble starts to migrate through the 
small bubble
network, faster
than the mean flow, when the foam velocity is large enough to induce a driving effect on the large bubble greater larger 
than the plastic threshold  of the foam  (see 
Fig. \ref{migra}).

Our main theoretical results consist of two scaling laws derived from general
dimensional arguments for the velocity threshold (eq. \ref{vth}) and for the large bubble velocity beyond that threshold (eq. \ref{anal_approx}).
   They are compared with the results of numerical simulations based on
the vertex model \cite{okuzono95} and with those of experiments 
presented in  \cite{cantat05b}.

The paper is organized as follows.
In Section \ref{2Dfoam}, we discuss the main ingredients (bubble
shape,
viscous force, etc.) that play a role in the process by which the 
instability takes
place. Section \ref{numerical} presents a general frame, in which the foam is
described by the positions of its vertices, that may be used for 
numerical simulations of any 2D foam flow.
Section \ref{qualitative} is devoted to the qualitative description 
of the performed simulations
and of the flow behavior so obtained. An analytical
model, based on a continuous description, is introduced in Section 
\ref{analytical}, and allows  quantitative comparison with the 
numerical results
described in
Section \ref{eqmotion}. The pressure  field in the
whole foam is discussed in Section \ref{pressurefield}. Finally, the 
results and perspectives are
summarized in Section \ref{summary}.

\section{Specific properties of the 2D foam.}
\label{2Dfoam}

\subsection{Equilibrium film shape.}
\label{equilibrium}

The two-dimensional foam referred to here is a foam
confined between to plates separated by a distance
  $h$, much smaller than the typical
bubble size.
For small values of $h$, the bubbles are organized in a single layer 
and every film touches
the upper and
the lower
plates \cite{cox02}. The
Plateau borders touching
the upper (or lower) plate  draw a network $\Gamma$ of
connected curved lines, as shown
on Fig.\ref{migra}.
  For 2D foams in equilibrium, the films separating two bubbles are part of vertical cylinders and
the only non vanishing curvature is thus the constant curvature in a 
plane parallel to the plates,
denoted by $c_1$ in the following. Thus $\Gamma$ contains the whole 
information on the foam
structure.

This property entails many important consequences
  specific to 2D foams in  equilibrium \cite{weaire} and yields a way of
measuring the bubble
   pressure by simple image analysis using the Laplace relation
$\Delta P =  \gamma c_1$, where $\Delta P$ is the pressure difference 
between two
adjacent
bubbles and $\gamma$ the surface tension.
In a flowing foam, the viscous forces, which are localized close to the plates,
modify the contact angle $\theta$ 
as depicted on Fig.\ref{force_plat} and \ref{courbure2}.
Both curvatures may be then of the same order of magnitude as evaluated in
section \ref{out_eq_shape} and the pressure
can no more be read on a foam image, as the mean curvature is not measurable easily anymore.

\subsection{Viscous forces.}
\label{par_visc}
The effective viscosity of a liquid foam pushed between two plates,
{\it ie} the
driving force to velocity ratio,
is typically one or two orders of magnitude larger than that of the pure liquid
phase,
despite the very large amount of gas ($>95\%$)  trapped in the material
\cite{schwartz87,hirasaki85,xu03,cantat04}.
This surprising behavior is due to the existence of very small length scales at
which the liquid
phase is confined in the structure. In the 2D geometry, the dominating process
is the force between
the Plateau borders and the plates. The viscous dissipation may have various
origins, including the bulk viscosity of the liquid soap water,  the 
viscosity of the surfactant monolayer
and the diffusion resistance, leading to different relations between
pressure drop and velocity \cite{buzza95,denkov05}.
  Considering a fluid surfactant monolayer, we assume here that the 
dominating contribution comes from the bulk viscosity. The case of a 
rigid monolayer
  will nevertheless be considered too in section \ref{par_seuil}.

The local flow geometry close to the plates is very similar to the
well-known Landau-Levitch situation, in which a plate is pulled out of a liquid
\cite{landau42,mysels}. The Plateau borders
play here the role of the liquid reservoir and the relative motion between the
liquid and the solid plate
is due to the motion of the bubble walls.
According to the Bretherton theory \cite{bretherton61}, the highest velocity
gradients occur in a small domain
connecting the Plateau borders and the wetting films, assumed to be at rest on
each plate.
This region is of characteristic thickness \cite{bretherton61}
\begin{equation}
\delta(v) \sim R_{Pl} (\eta_w v/\gamma)^{2/3} \sim 5 \, 10^{-7}m
\label{epaisseur_film}
\end{equation}
with $\eta_w = 10^{-3} Pa\, s$ the liquid phase viscosity,
$\gamma \sim 30 \, 10^{-3} N/m$ the surface tension, $v \sim 10^{-2} m/s$ the
macroscopic velocity of the Plateau border and $R_{Pl} \sim
10^{-4} m$ the Plateau border size.
The relevant Reynolds
  number for this problem is thus $Re = \rho \, \delta\, v / \eta_w \sim
5\,10^{-2}$, whose magnitude allows us to neglect the non-linear
convective term in
the Navier-Stokes equation.
The viscous force $\vec{f}_{v}^0$ (per unit length of Plateau border) exerted on a
  Plateau border sliding at a velocity $\vec{v} = v \vec{u}_v$ on the plate
  varies nevertheless non
linearly with $v$, namely as a power law with an exponent
  smaller than unity. It will be written $\vec{f}_{v}^0 = - \eta(v) \vec{v}$, $\eta(v)$ being an effective viscosity
 (with the dimension of a dynamic vioscosity).
  As the films swell when the foam velocity increases (see eq. \ref{epaisseur_film}), velocity 
variations occur on larger
  length scales and  the effective viscosity decreases.
  We assume here that $\eta(v) \sim 
v^{-1/3}$ (Landau Levich exponent), although power laws with exponent values ranging between
  $-1/3$ and $-1/2$ are reported in the literature \cite{denkov05}.
This force per unit length of Plateau border is finally of the order of $10^{-3}N/m$
and is expressed as
\cite{cantat04}
:
\begin{equation}
\vec{f}_{v}^0 = - \eta(v) \vec{v}  = -\lambda \eta_w v   Ca^{-1/3}
|\vec{u}_v \cdot \vec{n}| \vec{u}_v ; .
\label{fvisc_loc}
\end{equation}
with $Ca= \eta_w v / d$ the capillary number of the order of $3  \, 10^{-4}$.
The numerical prefactor $\lambda \sim 10$  and the geometrical factor were
measured in similar flow conditions, the latter
involving $\vec{u}_v$ the unit vector of the  film velocity direction and
$\vec{n}$ the unit vector normal to the film
(with an arbitrary orientation) \cite{cantat04}.
The force direction is not clearly evidenced and seems difficult to determine
directly.
It is here assumed to be in the direction opposite to that of the 
film velocity.

It will be useful to express instead the viscous force per unit surface of the 2D foam averaged on the bubble scale (mean viscous stress on the plates).
With a typical bubble size of $d \sim 10^{-2} m$ we obtain
\begin{equation}
{\bf F}_v = -{\lambda \eta_w v \over d}  Ca^{-1/3} \vec{u}_v\sim 
10^{-1}N/m^2 \; .
\label{Fvsurf}
\end{equation}
For comparison the viscous stress obtained for pure water with the same
mean velocity
is of the order of $\eta_w v / h \sim  10^{-2}N/m^2$.

\subsection{Out of equilibrium film shape}
\label{out_eq_shape}
\begin{figure}[h]
\centering
\includegraphics[width=5cm]{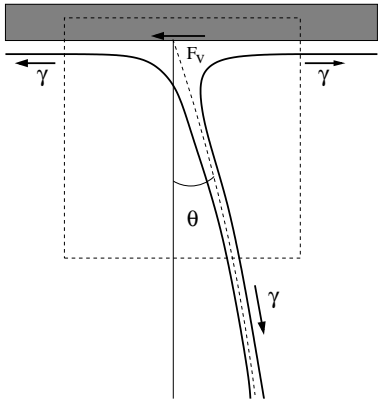}
\caption{\it A schematic side view of the Plateau border during motion.}
\label{force_plat}
\end{figure}
\begin{figure}
\centering
\includegraphics[width=5cm]{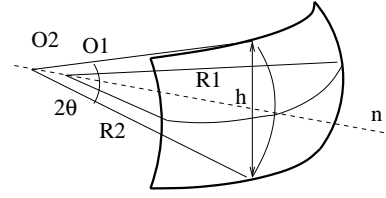}
\caption{\it The film curvatures. The Plateau border curvature is 
locally  $c_1= 1/R_1$. In the plane normal to the plates and to the 
Plateau border the film section can be approximated by an arc of 
circle of radius $R_2$.}
\label{courbure2}
\end{figure}

During flow, as inertia and dissipation in the
  bubble walls are negligible, bubbles still obey the Laplace equation 
$\Delta P =
\gamma c$ where c is the sum of the two main curvatures which remains 
constant on the whole bubble wall.
The boundary conditions on the plates are modified by the viscous
force, leading
to another film shape.
The force balance on an infinitesimal volume around a Plateau border (see Fig.
\ref{force_plat})
allows to determine the orientation
$\theta$ of the fluid film at the contact with the wall,
\begin{equation}
  \gamma \sin \theta =f_v^0 \; .
\end{equation}
The film shape is thus a surface of constant mean curvature with a 
contact angle
at the plates depending on the local velocity.
The two principal
curvatures vary on the surface but may be approximated by $c_{p1} 
\sim c_1=1/R_1$ and $c_{p2} \sim c_2=1/R_2$, with $c_1$ the curvature 
in the symmetry plane between the two plates  and  $c_2 = 2 \sin
\theta/ h \sim 30 m^{-1}$ (see Fig. \ref{courbure2}).
Both curvatures may thus be of the same order of magnitude
  for centimeter-sized bubbles. As the mean curvature remains constant 
on a given film,
if  $c_2$
varies along the film, then
$c_1$ varies too, and  the Plateau border is no more an arc of
circle.

\section{Numerical simulation of a 2D flow}
\label{numerical}

Simulations or
analytical calculations make a
compromise between the complexity of the dissipative function they use
\cite{schwartz87,reinelt89,li95}
and the number of bubbles they can consider 
\cite{glazier92,okuzono95,durian95,cantat03a,kern04}.
The numerical model presented in this section is well adapted for a broad
family of flow configurations in 2D.
It is based on a realistic dynamical behavior, even at large velocities, and on
a simplified  foam structure description
which allows to use a large number of bubbles at reasonable computational cost.

\subsection{Numerical variables and equation of motion}

As discussed above, the 3D film shapes are unknown.
A local description of the 2D Plateau borders network $\Gamma$  has been used
by Kern et al. in numerical simulations performed with
  a small number of bubbles in which only $c_1$ is taken into account  while
$c_2$  \cite{kern04} is neglected.
This approach requires
to discretize every Plateau border line and becomes very demanding 
for large scale simulations.
  For numerical efficiency, we choose to simplify the structure further
by assuming vertical films to be planar, as
already done in \cite{okuzono95,cantat03a}.
The Plateau borders  touching the plates are straight lines (called edges),
entirely determined by their end points
(the vertices).
Our numerical results have been  quantitatively reproduced by Sanyal 
et al. with
  a method based on a Potts model \cite{sanyal05}.

\begin{figure}
\centering
\includegraphics[width=6cm]{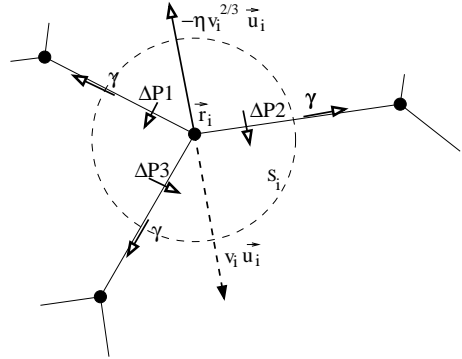}
\caption{\it Force balance on the system $S_i$ around the vertex $i$}
\label{sysnum}
\end{figure}

The variables used  are  the vertex locations ${\bf  r}_i$.
The three edges which meet at a given vertex i have their end points located at
${\bf  r}_i$ and ${\bf  r}_j$ with $j \in J_i$. The system $S_i$ is
then defined
as being composed of the vertex i with its three outgoing edges
reduced by half,
as depicted on Fig. \ref{sysnum}.
The equation of motion is provided by the force balance on each system $S_i$.
The number of vertices remains constant as we forbid the occurrence of film
breakage and of foam coarsening.

The tension force is, with ${\bf r}_{ij}={\bf r}_{i}-{\bf r}_{j}$ and 
$r_{ij}= || {\bf  r}_i-{\bf  r}_j ||$,
\begin{equation}
\vec{F}_{t,i}=\gamma h \sum_{j \in J_i}{ {\bf r}_{ij}  \over r_{ij}}
\label{Ft}
\end{equation}
We define $\vec{n}_{ij}$ to be the normal to the edge $(ij)$, oriented
arbitrarily, say from a bubble A
towards a bubble B, and $\delta P_{ij}(= P_B-P_A)$ to be the pressure
jump  on this
edge.
The resulting pressure force is then :
\begin{equation}
\vec{F}_{p,i} = - h \sum_{j \in J_i} {r_{ij}\over 2} \delta P_{ij} \vec{n}_{ij}
\label{Fp}
\end{equation}

Finally the viscous force determination imposes some additional approximations.
The whole system $S_i$ is assumed to  move with the same velocity as the vertex
(i) : ${\bf v}_i=v_i {\bf u}_i $.  Thus we get
\begin{equation}
\vec{F}_{v,i} = - k \sum_{j \in J_i}|{1 \over 2}{\bf  r}_{ij}  \times  {\bf
u}_{i} | v_i^{2/3}  {\bf  u}_{i}
\label{Fv}
\end{equation}
with the prefactor k obtained from eq. \ref{Fvsurf},
  $k= \lambda \eta_w^{2/3} \, \gamma^{1/3}$.

The equation of motion, $\vec{F}_{v,i}+\vec{F}_{t,i}+\vec{F}_{p,i}=0$ imposes
the orientation of $\vec{v}_i$ along
the unit vector $\vec{F}_i/F_i $, with $\vec{F}_i=\vec{F}_{t,i}+\vec{F}_{p,i}$.
We get thus the following explicit
expression for the velocity, as a function of the positions $\vec{r}_j$, and of $F_i$ which can itself be expressed as a function of the positions and the pressures (eq.\ref{Ft},\ref{Fp}) :

\begin{equation}
  \vec{v}_{i} =  \vec{F}_{i} F_i^2 \left (k \sum_{j \in J_i} |{1 \over 2} {\bf
r}_{ij} \times  {\bf  F}_i |  \right)^{-3/2}
  \label{eq_num}
\end{equation}
As we will see now, the pressures are also related to the positions, so that the velocity is a function of the only variables 
$\vec{r}_j$. An iterative process is then possible as, for each time step, the velocities and consequently the displacements of the vertices can be computed, leading to an actualisation of their positions.

\subsection{Explicit expression of the pressure}

On the time scale of a few seconds considered in this paper, gas diffusion is
negligible but pressure equilibrium
is reached. The pressure is thus uniform in each bubble $k$ and can be
determined, under the assumption of an ideal gas, as a function of 
the area $A_k$ of the bubble in contact
with the plate, knowing the constant quantity
of gas $n_k$ trapped inside and its isothermal compressibility 
$\chi_T = 1/P_0$ :
\begin{eqnarray}
P_k &= & {P^0 A_{k,0} \over A_k } \simeq P^0 - P^0\, { (A_k - A_{k,0}) \over
A_{k,0} } \; .
\label{PdeV}
\end{eqnarray}
where
\begin{equation}
A_{k,0}=  n_kR_{\small PG} T
/ h P^0
\end{equation}
is a reference value, obtained if the bubble
  pressure is equal to the global reference
pressure $P^0$. This pressure is the same for every bubble in the foam,
in contrast to the local reference $P_{eq}$ that will be introduced 
in the following
paragraphs.
In our simulations, the reference
pressure $P^0$ is strongly underestimated to
enhance the numerical stability. Nevertheless, even if $P^0$ is smaller than its experimental value, it remains much larger than 
all the pressure variations induced by the flow and the system is in the incompressible limit (the
relative variations of the bubble areas
remain smaller than few $\%$). The 
pressure field given by the product $P^0 \delta A/A$ is thus independent 
on $P^0$, even if both factors obviously depend on it.
The choice of the compressibility value (or equivalently  of the 
pressure reference) has therefore no real influence on the
dynamical
behavior, as checked numerically.

This formalism presents the big advantage to provide an explicit expression of
the pressure as a
  function of the bubble geometry, in contrast to the classical Lagrange
multiplier approach in which a
  strict incompressibility is assumed.

  The reference area is chosen for each bubble as a fixed parameter. The actual
area is computed after each displacement
  as a function of the positions of the vertices, assuming a
polyhedral shape for
the bubble. Finally, the pressure value in the bubble
  is deduced from eq. \ref{PdeV}.

Note that the Plateau rules which constrain the values of the 
angles between the tangents of connected films at vertices, which are 
for instance equal to
$120^o$ in a foam at equilibrium, are not imposed in this modeling 
because the films are
represented by straight lines. In this case, the pressure difference 
between adjacent bubbles
is no more related
  to a film curvature but to a deviation of the angles from the
reference value of
$120^o$.

\subsection{Building of an initial foam structure}

Periodic boundary conditions are imposed to the foam structure on the sides of
the rectangular simulation box, the direction of the flow being parallel to the
longer side.
First of all, an artificial network  of a few thousands of bubbles is
built, with
the positions and
  the connectivities of all vertices.
Most results were obtained with an initial structure built from a perfect
hexagonal network,
but few simulations
  were performed with a disordered foam obtained from a Voronoi tessellation.
Then a reference area is attributed
  to each bubble to reach the sought-after foam polydispersity and to create
possibly very large bubbles in the foam.
   The system ${\cal S}_0$ is then relaxed to an equilibrium foam 
structure, using  the
   iterative process detailed in paragraph \ref{temporal}
   until the vertex motions become of the order of the numerical 
noise. After this first relaxation step, the
areas are very close to the reference areas $A_{k,0}$.

\subsection{Temporal evolution}
\label{temporal}

\begin{figure}
\centering
\includegraphics[width=5cm]{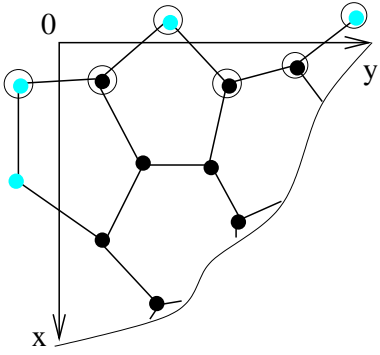}
\caption{\it The encircled vertices belong to the line $\cal{L}$.
Their velocities are fixed to a given value to impose the flow rate.
The two axes are sides of the simulation box, gray vertices result from the use
of  periodic boundary conditions.
}
\label{boundary}
\end{figure}

The foam evolution is based on eq.\ref{eq_num}. It is determined iteratively in
time from an initial structure ${\cal S}_0$, that may
  be either an equilibrium shape or an arbitrarily chosen out-of-equilibrium
shape.
  Both transient or stationary processes may then be studied
either in the presence of external perturbations or during relaxation to
equilibrium.
If we want to impose a flow in the direction x, we choose a set of vertices
connected by a wiggly line $\cal{L}$
of edges
whose average orientation is along y (Fig. \ref{boundary}). For that 
purpose, we select first a vertex with a first neighbor on the other 
side of the line
$y=0$. Then, from 
this point $\bf{r}_i$, we select the neighbor of smallest $|x|$ 
verifying $y>y_i$, and we iterate the process until we recover
the initial point.
   The motion of the vertices which belong to that
set $\cal{L}$ does not obey eq. \ref{eq_num} as do the other vertices,
but they are
forced to move at the desired velocity at each time step.
This induces an artificial pressure discontinuity across the line $\cal{L}$,
connecting the upstream and the downstream regions in the periodic box.
The information propagates through the whole foam thanks to the pressure field.

The main algorithm is the following :
During the time step $dt$,
(a) the vertices belonging to $\cal{L}$ are displaced from ${\bf v}_0 
dt$(in case of flow).
(b) the bubble areas and pressures are updated.
(c) Velocities, and thus displacements, are computed everywhere else 
with eq.\ref{eq_num}.
(d) If the vertex displacements result in the formation of too small edges, T1
events are performed (see below), otherwise,
(d') the displacements $\vec{v} dt$ are performed.

\begin{figure}
\centering
\includegraphics[width=5cm]{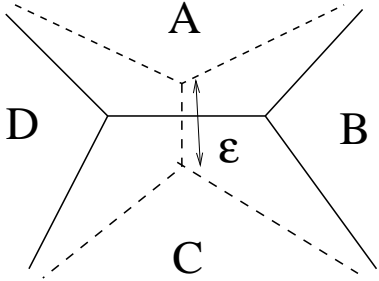}
\caption{\it The structure before (dashed lines) and after (solid lines) a T1
transformation.
}
\label{T1}
\end{figure}
If an edge becomes smaller than a fixed value $\epsilon$, a topological
transformation T1 is performed, as explained on Fig. \ref{T1}.
For stability reasons, only one T1 is allowed to occur at each time 
step in the whole foam. In the rare case where
  two edges become simultaneously too small, we simply choose to 
perform only one
of the two transformations,  the other being done at the next time 
step. This would have to  be improved to deal with different
situations, for instance the case of foam shearing.
The edge is switched perpendicularly to its previous orientation, and get a
new length $\epsilon' \sim \epsilon$, which is an arbitrary 
constant parameter,  and the connectivity of
  the four neighboring bubbles is modified. This transformation induces a
concentration of tension forces around this new edge
  that leads to a rapid
  local relaxation governed by the same dynamical equations as the 
rest of the motion.
As the process is very local, the 2D description presumably fails and 
the dissipation in the
two vertical Plateau borders involved in the T1 must be of the order
of the dissipation on the plates. This additional dissipation will be 
taken into account in a
future modeling.  In any case, if the characteristic time of a 
relaxation after a T1 remains small in comparison with the
other time scales, they can be seen as instantaneous processes and 
their specific dissipation rates
do not modify the foam behavior.

\section{Large bubble instability : qualitative behavior}
\label{qualitative}
\subsection{Performed simulations}

We investigated the behavior of a monodisperse foam in which a single large
bubble has
been created. This defect is produced during the first relaxation step by
imposing a large
reference area to an arbitrary bubble chosen upstream, called LB. The 
results presented below
were obtained with a LB area
20 times larger than the small bubbles area. The mean flow velocity is 
the control parameter
of our study.
Once the foam is equilibrated, 
 the flow is turned on. The bubble areas adjust then 
slightly in order
to establish the
macroscopic pressure gradient between upstream and downstream. In the frame of
the mean flow (in which $\cal{L}$ is at rest), a small upstream 
motion of the bubbles
is then observed, corresponding to an area decrease upstream and an area
increase downstream.
Simultaneously, the foam deforms around the large bubble which tends to move
faster than the mean flow.
If the mean velocity is
smaller than a threshold value, a stationary shape is obtained and the flow
remains a plug flow.
Otherwise, T1's occur around LB which begins to migrate.

The parameters $\epsilon$, $\epsilon'$, $h \,P^0$, the box size and 
the time step
were varied
to test the numerical stability. The algorithm becomes  numerically 
unstable for a too large time step, whose maximal acceptable value
is mainly related to $h \,P^0$  and to $\epsilon$. The stability 
domain has not been systematically investigated,
but we carefully checked the reproducibility of our results over a 
large range of parameters.
Only $\epsilon$ has a noticeable influence on
the threshold value as physically expected.
It may indeed be seen as the characteristic size of a
   Plateau border, and is thus related to the liquid fraction  $\phi$ 
of the foam by the relation
   \begin{equation}
   \phi \sim \epsilon^2 d / ( h d^2) \sim  \epsilon^2 / (d h )  \, ,
   \label{fract_liq}
   \end{equation}
   with $d$ the typical edge length, $\epsilon^2$ the Plateau border 
section area  and h the distance
   between the plates.
Except for the results shown in Fig. \ref{epsilon}, all results were obtained
with a value of  $\epsilon$   which
corresponds to a liquid fraction of $0.5 \%$ (disregarding the additional water
layer wetting the plates).

\subsection{Disordered and ordered foams}

\begin{figure}[h]
\centering
\includegraphics[width=7cm,angle=-90]{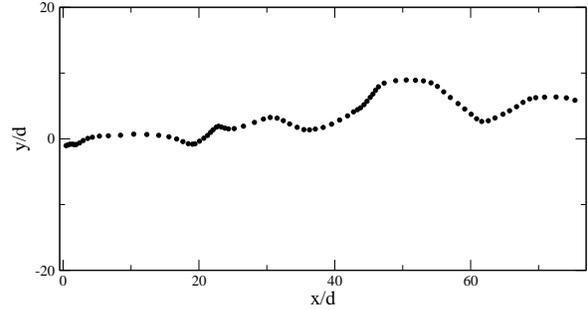}
\caption{\it Trajectory of the large bubble center of mass in the 
mean flow frame in a case of a disordered foam (the flow is
in the x direction). The distance between
two successive points is proportional to its relative velocity.
}
\label{traject}
\end{figure}

In a monodisperse disordered foam, we don't observe a stationary motion of the
large bubble for a velocity larger than the threshold.
The shape of the large bubble, the magnitude and the direction of its 
relative velocity 
fluctuate strongly (Fig \ref{traject}).
When the large bubble moves at its highest velocity, its shape is elongated in
the direction of its relative motion.
The pinning of the large bubble by some defect in
the disordered foam stops its migration. The bubble shape becomes
then elongated
in a direction perpendicular to that of
the mean flow. Finally this shape destabilizes, by developing a tip downstream.
This tip grows by producing few T1's
in the front of the large bubble and finally absorbs the full bubble and a new
cycle begins.
This qualitative behavior is in good agreement
  with experiments \cite{cantat05b}. This jerky motion is puzzling and its
extensive study
  is the aim of a future work.
The explanation of the whole instability process requires first to understand
the mean behavior of the large bubble. This is the aim
of the present work.

The following results were thus obtained
with an hexagonal network, only distorted by a single defect.
A stationary regime, in which the large bubble moves at constant 
velocity,  is reached after a very short transient.
Despite the overall homogeneity of 
the foam, various kinds of motions may be
observed
with various stabilities and probabilities of occurrence.
In the most frequent situation, the large bubble migrates solely thanks to T1's
that take place
near its front and its rear. The motion is periodic, and the large bubble
recovers exactly the same
neighborhood after a migration of one bubble layer.
Only few T1's, involving the first or second LB neighbors, are performed during
this period.
The crystalline organization is perfectly restored beyond the large bubble.
The orientation of the relative LB motion may differ strongly from
the direction
of the mean flow.
It occurs always along the most favorable crystalline direction (see
Fig.\ref{migra}) and
the angle between the mean flow and
the relative velocity of the large bubble may thus reach $30^o$.

More rarely, for slightly different initial conditions,
  a well defined wake may appear, with few bubbles moving behind the 
large bubble with the same
velocity.
  This is observed
in experimental flows too.
Several choices seem thus to be offered to the large bubble, which may explain
the large fluctuations  observed
experimentally and numerically in disordered foams.

Detailed numerical results found for the velocity threshold and the relative
velocity of the
large bubble are presented in the next section together with analytical
predictions.
They are obtained for LB trajectories in the direction
of the flow, without wake, which is the most probable situation.
   The velocities, averaged over a set of
experiments, are in fair agreement with predicted values as shown on 
Fig.\ref{vit_expe}.

\section{Continuous description}
\label{analytical}

The numerical results are compared hereafter to analytical 
predictions based on a
completely different point of view.
We partially forget the specific foam structure, at least outside  the large
bubble, and we treat it
as an elasto-plastic continuum medium. This point of view has already 
been successfully adopted
by Kabla et al. in a quasi-static context \cite{kabla03}.
  Scaling behaviors are determined to emphasize the main physical processes
involved, independently of
any numerical prefactor.
We give in the following subsections the analytical expressions of the various
forces exerted on the
network of small bubbles, per unit surface of foam (as seen from above),
averaged over a few bubbles.

\begin{figure}
\centering
\includegraphics[width=6cm]{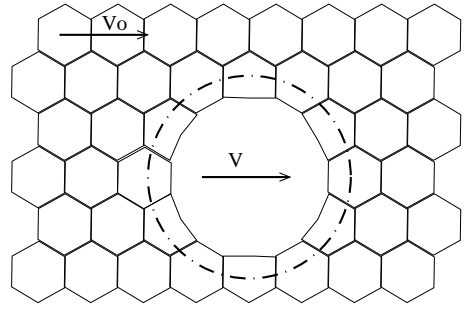}
\caption{\it Internal system and external system, separated
by the dashed circle.
}
\label{systeme}
\end{figure}

\subsection{Viscous forces}

Below the threshold, the viscous forces are well determined, as the velocity
field is ${\bf v}(x,y)=v_0 {\bf u}_x$
everywhere.
The migration of the large bubble induces of course velocity 
variations in the foam, but
the LB may move at high speed
among the others with only very localized variations of vertices (and 
films) velocities. The
migration occurs through a fracture
process : few vertices at the front of the large bubble are 
accelerated until they
reach their downstream neighbors
and undergo a T1.  Then the small bubbles initially in contact with 
these vertices
move on the sides of LB and their
vertices recover the mean velocity $v_0$. The vertices between LB
and their new
downstream neighbors
are then accelerated at their turn.
A similar but inverted transformation occurs behind the large bubble.
Thus velocities higher than $v_0$ and T1 transformations are confined in the
internal system defined on Fig.\ref{systeme}.
In the external system, the variations of the network deformation due to the LB
migration occur
without topological rearrangements and induce only negligible velocity
gradients.

For these reasons the velocity field is  assumed to be $v_0  {\bf u}_x$  in the
whole external
domain, even above the threshold.
The viscous force per unit foam surface is thus (see
eqs.\ref{fvisc_loc},\ref{Fvsurf}),
\begin{equation}
F_{visc} = -{\eta(v_0) v_0 \over d}{\bf u}_x=-\nabla \left ({ \eta(v_0) v_0 x
\over d } \right )
\label{eq_f_visc}
\end{equation}

\subsection{Pressure field}

The pressure in a foam is heterogeneous, even at equilibrium. Its value is
mainly related to the foam topology and to
the number of sides $n$ of each bubble and arise from the
competition
between tension forces which
tend to shrink the bubbles and pressure
forces which tend to inflate them. This effect remains present
even when the
foam flows and is then superimposed with purely dynamical pressure variations.
To better evidence the dynamical contributions, we define the equilibrium
part of the pressure in a given flowing structure as the pressure
in each bubble after relaxation to the closest equilibrium shape.
The latter shape is obtained
numerically by fixing the line $\cal{L}$ of vertices and relaxing the rest of
the foam.
  As the foam is only locally deformed, relaxation
occurs usually, as desired, with no need of T1 transformation.
This somewhat artificial splitting of the total pressure in two terms proved to
be very powerful, as it allows us to reach a full and simple
analytical description of the complex pressure field, in good
agreement with the results of computer simulations.
  Hereafter, the equilibrium pressure defined in that way will thus be 
the local reference
pressure, remembered to be specific to each bubble.
Its value is discussed in detail in paragraph \ref{pressurefield}.
Our method  should however be improved to tackle more complex 
situations in which the
system can only relax through many T1 transformations.

The resulting force per unit foam surface due to the heterogeneous
gas pressure
is consistently split into two terms, namely:
\begin{equation}
{\bf F}_{p} ={\bf F}_{p,eq} - h {\bf \nabla} \bar{P},
\label{pressure2}
\end{equation}
The first term of the right-hand side is by definition obtained in a 
foam at equilibrium, as discussed
above. It will
play no role in the dynamics.
The second term is not very sensitive to the local structure
of the foam, and we assume that it is correctly described by the gradient of a
smooth dynamical pressure
field defined as  $\bar{P}=P-P_{eq}$.
This field will be derived in section \ref{pressurefield}.

\subsection{Tension forces}

Liquid foams have a rather simple behavior at small strain.
A linear elastic response is obtained in this case with an elastic coefficient
scaling as
$\gamma/d$ \cite{weaire94,hohler97}. If the stress
reaches a given
threshold, equally of the order of $\gamma/d$, plastic deformation
occurs (through T1's).

The distribution of surface forces, due to tension forces, is first decomposed
in the same way as the pressure, leading
to
\begin{equation}
{\bf F}_{t} = {\bf F}_{t,eq} + {\bf \bar{F}}_{t}
\label{tension}
\end{equation}
with ${\bf F}_{t,eq}= -{\bf F}_{p,eq}$.

Considering an elastic and incompressible response of the foam we get
\begin{equation}
{\bf F}_{t} = -{\bf F}_{p,eq} + {\gamma h \over d} \nabla^2 {\bf X} \; ,
\label{tension2}
\end{equation}
where $X$ is the displacement vector from the closest equilibrium shape
obtained after relaxing the foam as explained previously. The weak foam
compressibility is disregarded.

\section{Instability threshold and large bubble velocity}
\label{eqmotion}

\subsection{Driving force}

Using the three force expressions eqs.
\ref{eq_f_visc},\ref{pressure2},\ref{tension2} and neglecting any
inertial term,
we get the following
equation for the external system,
  \begin{equation}
     -\nabla \left ({ \eta(v_0) v_0 x \over d }+ h \bar{P} \right ) + {\gamma h
\over d} \nabla^2 {\bf X}
    =0   \; .
    \label{eqmot}
    \end{equation}
The equation of incompressibility div X=0 and the
boundary conditions must be added to eq. \ref{eqmot}.

The internal system shape
is approximated by a disc $\cal{D}$ of diameter D centered at the LB position.
As it tends to migrate towards positive x, its displacement is 
$\delta \vec{u}_x$.
The parameter $\delta$ remains
undetermined at that point, it  will be
expressed explicitly in the next section.
So we get, as first boundary condition for eq. \ref{eqmot},
   ${\bf X}= \delta \vec{u}_x$ on $\cal{D}$.
In computer simulations, boundary effects on the cell sides are 
actually disregarded as the
problem is solved with periodic boundary conditions.
In contrast, in real experiments in a Hele-Shaw cell of width $2 L$ there is a
frictionless slip on the lateral smooth boundaries, leading  to  $X_y= 0$ and
$\sigma_{xy}=0$ where $\sigma$ is the elastic stress tensor.
Boundary conditions have only a logarithmic influence on the force 
fields near the
large bubble, varying as $ln(D/L)$. They are difficult to determine
analytically
and are not relevant for our purpose.
  As an illustration, we nevertheless
give here the asymptotic expression of the elastic force exerted on the large
bubble
with a third condition, ${\bf X}=0$ on the lateral boundaries,  well documented
in the literature in the context of viscous hydrodynamics
\cite{faxen38,happel}.

The large bubble motion is governed by the resulting force induced by the
external pressure field and the
elastic stress acting on the internal system.
This force $F_x$, oriented along the mean flow for symmetry reasons is, noting
$\cal{C}$ the boundary of the internal system and $\vec{n}$ its normal,
\begin{equation}
F_x = h \int_{\cal C} \! - \bar{P} n_x \! + \! {\gamma \over d} \left ( 2
\drp{X_x}{x} n_x + \left(\drp{X_y}{x} + \drp{X_x}{y}\right) n_y \right ) dl
\label{drivingforce}
\end{equation}
The analytical expression of this force has been obtained by analogy with the
problem of an infinite cylinder pushed at constant velocity in a 
viscous fluid between two walls
which is
solved in the literature \cite{faxen38}. The identification of each
variable is based on the similarity between the equations governing 
an elastic solid and a viscous fluid. As detailed in Appendix 
\ref{analogie_hydro} it leads to
\begin{equation}
F_x = {\eta(v_0) v_0 D^2 \over d }-{\gamma h \over d}{4 \pi  \delta  \over
\mbox{ln}({2L \over D}) - 0.91 + 0.43 \, {D^2 \over L^2} +...}
\label{drivingforce2}
\end{equation}
The first term, oriented downstream, is the driving force for the instability.
It is
  a pressure contribution corresponding to the missing viscous forces in the
large bubble. It is similar to an Archimedes force in which the homogeneous
gravity field would be replaced by the
homogeneous viscous force field $- \eta(v_0) v_0/d \vec{u}_x$. The resulting
pressure force is proportional
to the force field intensity times the area of the hole.
The second contribution, oriented upstream, is the elastic response
of the foam.
As expected, it is proportional to the elastic
coefficient $\gamma/d$ and to $\delta$.
The denominator is an infinite sum expanded in the small parameter
$D/L$. Its
logarithmic dominating term is specific
to the 2D case, in which elastic problems cannot be solved without boundary
conditions at finite distance.
This explicit analytical solution allows to quantify roughly the influence of
the cell size. In any case, as other numerical prefactors
will be disregarded in the following, we neglect this logarithmic
dependence and
we only checked that the denominator
remains close to unity.
The driving force expression used in the next subsection is thus
\begin{equation}
F_x = {\eta(v_0) v_0  D^2 \over d }-{ \gamma h \delta  \over d}
\label{drivingforce3}
\end{equation}
 From this force, compared to the viscous force exerted on the large bubble
itself, we determine the
value of $\delta$ in the following.

\subsection{Instability threshold}
\label{par_seuil}
At small velocity, the elastic response of the foam compensates the
driving force
and the large bubble has the same velocity $v_0$ as the rest of the foam.
The viscous force applied to the internal system, {\it ie} that due to the foam
films around LB, is
given by $D \eta(v_0) v_0$ (see eq.\ref{fvisc_loc}). This  leads to 
the following equation for $\delta$,
with the use of eq. \ref{drivingforce3}:

\begin{equation}
{\eta(v_0) v_0  D^2 \over d }-{ \gamma h \delta  \over d} =  D \eta(v_0) v_0 \;
,
\label{eq_delta}
\end{equation}
and, in the limit $D\gg d$,
\begin{equation}
\delta \simeq { \eta(v_0) v_0 D^2 \over \gamma h }
\end{equation}

The threshold is reached when the maximal stress in the foam reaches the yield
stress $\gamma/d$.
The largest elastic stresses are localized in front of the large
bubble and just
behind
and their values are determined from dimensional arguments.
As the elastic coefficient scales as $\gamma /d$, they scale as 
$\gamma \delta / (d L_{geo}) $, where $L_{geo}$ is a length
related to the problem geometry. The box size appears only through logarithmic
corrections.
In a purely continuous medium description,  the small bubble size  scales
out and thus $L_{geo} \sim D$, which is the value retained
in the next part \ref{lbrv}.
Anyway, the discrete nature of the foam should probably be taken into account
here.
The small length scale $d$ influences the radius of curvature of the LB in the
region surrounding
the fracture tip, and therefore the stress concentration.
Both assumptions  $L_{geo}= D \mbox{ or } d$ are thus investigated below, as
each might be valid in some domain which depends on the actual aspect ratio.

 From this, we deduce the maximal value $\delta_{max} \sim L_{geo} $ and the
velocity threshold $v_{th}$:

\begin{equation}
\eta(v_{th}) v_{th} \sim \gamma {h L_{geo} \over D^2}
\label{etavth}
\end{equation}
Using  $\eta(v_{th}) v_{th}= \gamma (\eta_w v/\gamma)^{\alpha}$ (see eq.
\ref{fvisc_loc}) we get,
with the adimensional capillary number $Ca=\eta_w v/\gamma$,

\begin{equation}
  Ca_{th}  \sim  \left ({h L_{geo} \over D^2}\right ) ^{1 \over \alpha}
  \label{vth}
\end{equation}

\begin{figure}[h]
\centering
\includegraphics[width=6.5cm,angle=-90]{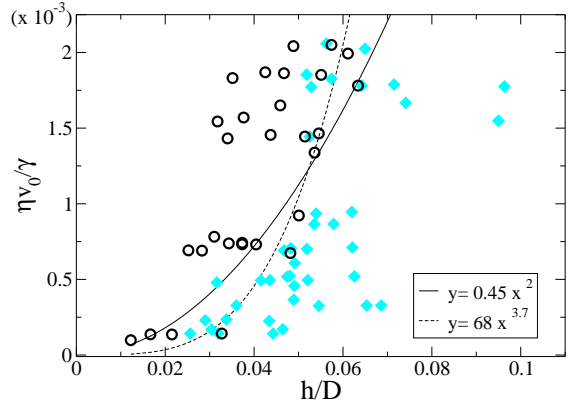}
\caption{{\it Experimental results from Cantat et al. \cite{cantat05b}. Each
point in the plane $(h/D ; \eta v_0 /\gamma)$ represents
a flow, with a large bubble migrating (circle) or not (square). Two domains
appear, separated by a boundary correctly fitted
by the relation $Ca_{th} = 68 (h/D)^{3.7}$(dashed line). Fluctuations 
produce an overlap
between the two domains which results in a large uncertainty
on the exponent value.  The boundary line, obtained from eq.\ref{vth} with
$L_{geo}=D$ and $1/\alpha = 2$ and an adjustable multiplicative
prefactor, is further plotted (full line).
}}
\label{seuil_expe}
\end{figure}

  Depending on the value of $L_{geo}$ and on the
value of $\alpha$ (between 0.5 and 0.66 as discussed  in section
\ref{par_visc}), the model predicts
an exponent for $D$ ranging between $-1.5$ and $-4$. The experimental 
value is of
the order of $-3.7$, in a case where $\alpha = 0.5$
(see Fig.\ref{seuil_expe}, from \cite{cantat05b}).
The choice $L_{geo}=d$, which is thus in better agreement with the 
experiments than
is $L_{geo}=D$, reinforces the importance of the discrete
structure of the foam. New
  numerical simulations, especially with a large range of large bubble size, are nevertheless needed to be fully conclusive.

\begin{figure}[h]
\centering
\includegraphics[width=6.5cm,angle=-90]{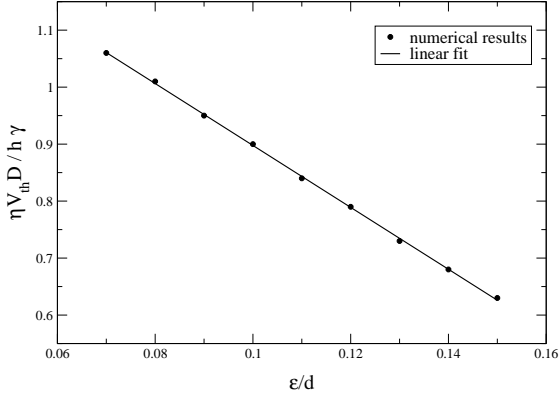}
\caption{{\it Velocity threshold as a function of the minimal 
distance $\epsilon$
  allowed between two vertices before a T1 occurs.
}}
\label{epsilon}
\end{figure}

The liquid fraction plays an important role in the plastic threshold
\cite{weaire94}, and therefore influences
the prefactor in eq.\ref{vth}.
In our vertex model simulation, the liquid fraction is related to the 
minimal distance
$\epsilon$  allowed between two vertices before a T1 occurs (see eq. 
\ref{fract_liq}).
The variation of the threshold with this parameter is displayed on 
Fig. \ref{epsilon}. All the
simulation results presented
on the other graphs were performed with a fixed $\epsilon$ value, as small as
allowed by the numerical
stability.

\subsection{Large bubble relative velocity}
\label{lbrv}
\begin{figure}
\centering
\includegraphics[width=6cm,angle=-90]{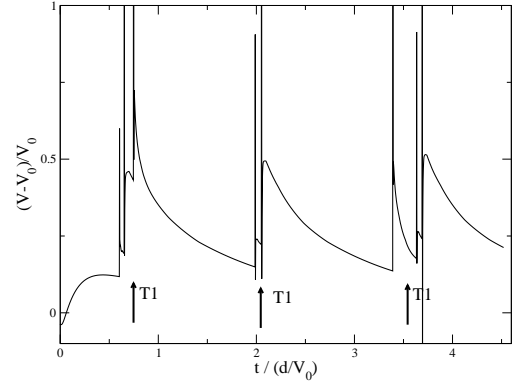}
\caption{\it Numerical LB velocity as a function of time, for $v_0 > v_{th}$.
T1's involving LB edges occur in avalanches of 2 or 3 events and are 
indicated by arrows.
  At these times the LB is suddenly deformed and the center of mass 
velocity diverges numerically.
  Just after a T1 avalanche, the
elastic stress is
strongly  reduced
and the
LB velocity is large. As the stress accumulates in the small bubbles 
network the
LB velocity decreases until the yield stress
is reached and new plastic events occur.
In contrast, if $v_0 < v_{th}$ the LB relative velocity tends to zero 
before the yield
stress is reached.
}
\label{vdet}
\end{figure}

For flow velocities larger than the threshold, plastic transformations occurs
regularly around the
large bubble which migrates through the foam with a velocity v(t).
As these T1's are strongly localized around LB, they can be 
considered as discrete
events relaxing
suddenly the largest part of the elastic stress. Our model for the LB behavior
  is thus the
following :
if a T1 occurs at time $t=0$, the elastic force vanishes, the value of
$\delta$ is zero and
the large bubble velocity is given by eq.\ref{eq_delta} :
$\eta(v(0))v(0) =\eta(v_0)v_0 D/d$.
Then $\delta$ increases with time, inducing  an elastic stress increase and
a LB velocity decrease. Finally $\delta$ reaches its limit $D$ at $t=t_{max}$
and a new cycle begins (see Fig. \ref{vdet}).

As $\delta(t)=\int_0^t (v-v_0) dt$ the equation of motion becomes, 
between $t=0$
and $t=t_{max}$,
(see eq.\ref{eq_delta})
\begin{equation}
- \gamma h /d \int_0^t ( v-v_0) dt + {\eta(v_0) v_0 D^2  \over d }= 
D \eta(v(t)) v(t)
\label{eq_mouv}
\end{equation}

\begin{figure}[h]
\centering
\includegraphics[width=6cm]{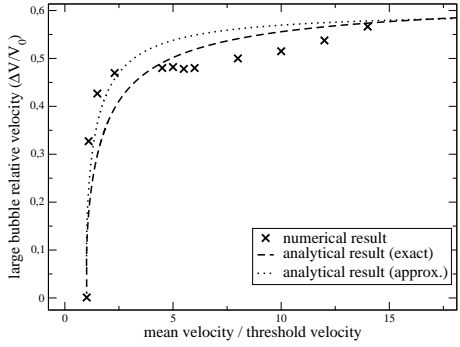}
\caption{{\it Large bubble relative velocity as a function of the mean flow
velocity. The first analytical law is given
by eq. \ref{anal_exact} and the second one comes from eq. \ref{anal_approx}.
Numerical prefactors
are adjusted to obtain the best fit between numerical and analytical values.
Their values are respectively 0.25 and
0.2, which is of the order of unity as expected. The ratio $D/d$ is 3.
}}
\label{vdev0fit}
\end{figure}

As detailed in Appendix \ref{sol_exacte}, this differential equation 
is analytically solvable.
The approximate
solution detailed below is obtained by neglecting the variation of $\eta(v)$
with $v$. It differs only by few percents from the exact one
and is much simpler (see Fig. \ref{vdev0fit}). Taking the time derivative, we
get

\begin{figure}[h]
\centering
\includegraphics[width=6cm,angle=-90]{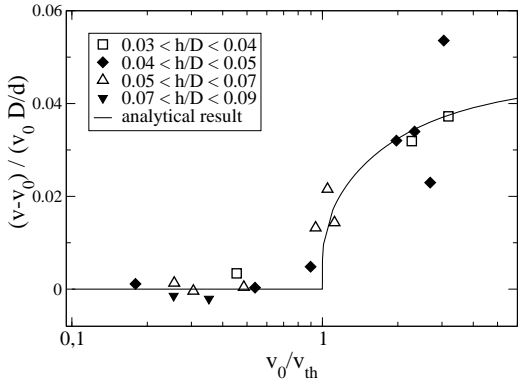}
\caption{{\it Experimental results from Cantat et al. \cite{cantat05b}.
Large bubble relative velocity as a function of the mean flow
velocity, rescaled by the threshold velocity obtained on 
Fig.\ref{seuil_expe}, for various large bubble sizes.
The analytical fit comes from eq. \ref{anal_approx}, with a single 
adjustable prefactor.
}}
\label{vit_expe}
\end{figure}

\begin{equation}
- {\gamma h \over d D \eta} \left (v(t) - v_0 \right ) =   {dv \over dt}
\end{equation}
\begin{eqnarray}
  v(t) &=&  v_0 +  (v(0)-v_0) e^{-t/\tau} \label{vdetexp}\\
   \tau &=& {d \,  D \, \eta \over \gamma\, h}
\end{eqnarray}

Expression \ref{vdetexp} is valid until $\delta$ reaches
$\delta_{max}\sim D$ at
time $t_{max}$.

\begin{equation}
\int_0^{t_{max}} \!(v(t)-v_0) dt = D = -\tau (v(0)-v_0) \left(
e^{-t_{max}/\tau}
-1  \right)
\end{equation}
So
\begin{equation}
t_{max}= -\tau \, ln \left( 1- {d \over \tau v_0 } \right)
\end{equation}
and finally, with $<>$ denoting the average in time,
\begin{eqnarray}
<v-v_0>&=& {D \over t_{max}}= {-D   \over   \tau ln \left( 1- {d \over \tau v_0
} \right) } \\
{<v-v_0> \over v_0 D/d} &=& {-v_{th} / v_0 \over ln \left( 1- {v_{th} \over
v_0}\right )}
\label{anal_approx}
\end{eqnarray}
The velocity threshold is consistently obtained from eq. \ref{etavth} with
$\eta=\mbox{const.}$,
  namely $v_{th}=\gamma h/\eta D$.

  This expression is in good agreement with our numerical simulations (see Fig.
\ref{vdev0fit}).
The small oscillations of the numerical points on
both sides of
the analytical curve are related to a
bubble structure modification around LB at some velocities.
The logarithmic behavior above
  the threshold renders the transition almost discontinuous. A small
modification
of the threshold value,
  due to a small polydispersity for example, induces a large variation of the LB
velocity.
  The fluctuations observed experimentally are therefore relatively important,
but the mean velocities, averaged
  over many experiments, are very well fitted by the previous law as 
shown on Fig.\ref{vit_expe},
  from Cantat and Delannay \cite{cantat05b}.

  A better understanding of fluctuations is beyond the scope of this article. It
would necessitate
  an improved description of the force field in the foam and of the structural
disorder.

\section{Pressure  field.}
\label{pressurefield}

\subsection{Pressure at equilibrium}

\begin{figure}
\vspace*{0.5cm}
\centering
\includegraphics[width=5.5cm,angle=-90]{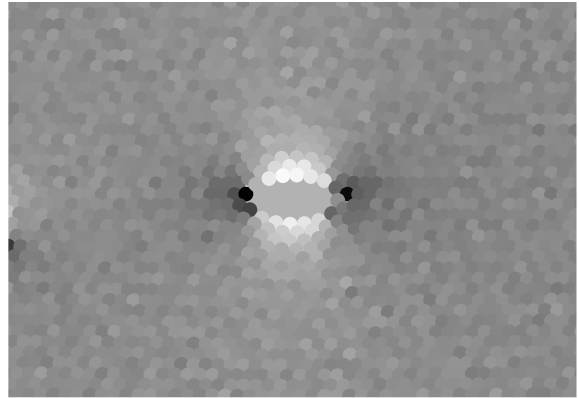}
\includegraphics[width=6.5cm,angle=-90]{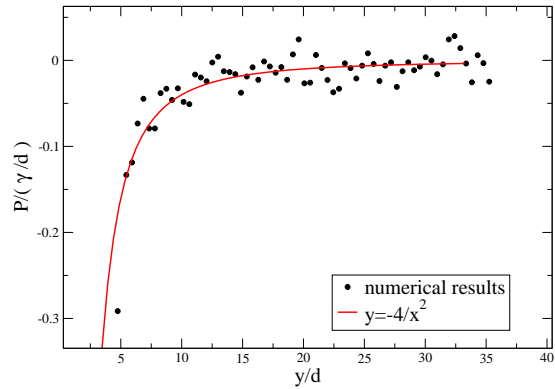}
\caption{\it Pressure field around the large bubble, obtained by
relaxation from
an out of equilibrium situation, with a velocity larger than the
threshold value.
(a) Pressure field in gray level : pressure decreases from white to black (for
clarity, the color of LB is arbitrarily chosen, its real pressure being lower
than that of small bubbles). (b) Pressure along the horizontal 
segment $y=0$ from LB
to the right boundary. The solid line is a fit to the analytical 
expression given by
eq.\ref{pmoinsp0}, with an adjustable prefactor. }
\label{presseq}
\end{figure}

The pressure at equilibrium in a 2D foam obeys elegant and simple 
rules \cite{weaire}.
It is mainly related to the number of sides of each bubble.
The average number of sides is six as a
consequence of Euler's relation in 2D \cite{weaire}.
The pressure inside bubbles with $n>6$ is larger than the external reference
pressure whereas it is smaller when $n<6$. When bubbles with $n \neq 6$ can be
considered
  as relatively isolated defects in a monodisperse foam, the pressure
distribution has been calculated by Graner et al. from an electrostatic
analogy. Bubbles with $n \neq 6$
play the role of topological charges and  modify the pressure field in the same
way as positive or negative charges produce an electric potential.
The induced pressure field around each defect is, with r the distance to the
defect,  (see \cite{graner01})
\begin{equation}
P-P^0= -{\gamma \over d } {(6 -n)\pi \over 3}  ln({r \over d})
\end{equation}
The logarithmic divergence is specific to the 2D geometry and we recover an
expression similar to the
electric potential produced by an infinite and uniformly charged line.
The various contributions are additive, and a multipolar expansion can thus be
used to derive
the pressure field far from the defects.
  During its migration, the large bubble
adopts an elongated shape which is kept even after relaxation. The distribution
of 5-sided bubbles on both sides of the large bubble with $n_{LB}$ sides is
equivalent
to a quadrupolar distribution of topological charges, from which we deduce the
dominant term in the pressure variation, up to a numerical prefactor,
\begin{equation}
(P-P^0)  \sim n_{LB} d \gamma \left ({2 y^2 \over (x^2 +y^2 )^2} - 
{x^2 \over (x^2
+y^2 )^2} \right ) \, .
\label{pmoinsp0}
\end{equation}
The results of numerical simulations are in good agreement with the predictions
of eq. \ref{pmoinsp0} as shown in figure Fig.\ref{presseq}a-b.

\subsection{Pressure during the flow}

\begin{figure}[h]
\centering
\includegraphics[width=5.5cm,angle=-90]{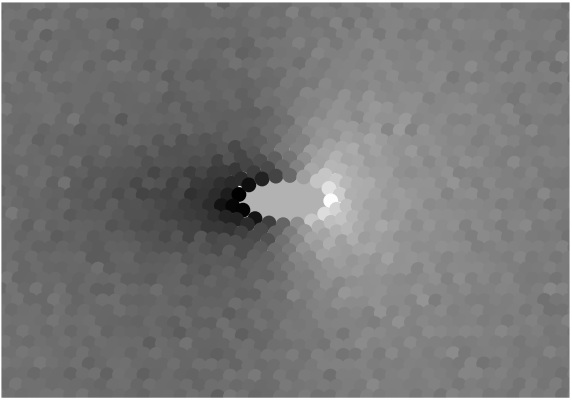}
\includegraphics[width=7cm,angle=-90]{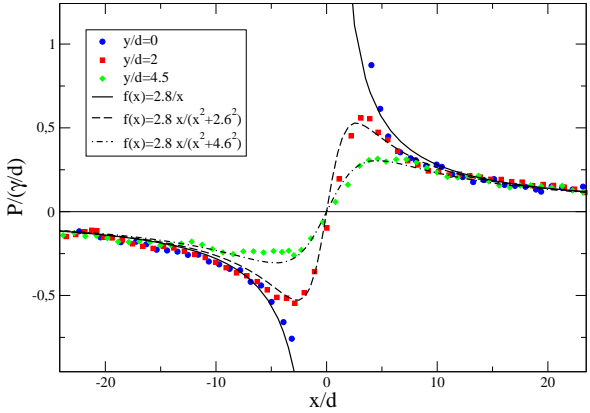}
\caption{\it Pressure field in the foam after removing the equilibrium and
linear contributions. The foam flows from the left $(x<0)$ to the
right $(x>0)$.
(a) Numerical results. Light gray corresponds to
high pressure values. (b) Another view of the same results :
pressure in the foam along the line $y=const.$, for three value of y.
The line y=0 goes from upstream to downstream, trough the middle of 
LB. The values are well fitted by the analytical expression 
\ref{barP}.}
\label{pressureheq}
\end{figure}

The pressure field in the flowing foam has already been discussed in detail in
\cite{cantat03a}, and we  just recall the main results here.
As it is less sensitive to long range effects than the displacement field, it
can be computed
analytically without taking into account the positions of the boundaries.
The final equation for the pressure is
, with $\vec{r}_0$ the position of the large bubble,
  \begin{equation}
     \bar{P} = - {\eta v_0 x \over d} + {\eta v_0  D^2 \over 2\pi d } {x -x_0
\over ({\bf r}- {\bf r}_0)^2} \, ,
     \label{barP}
     \end{equation}
The first term is the linear pressure variation responsible for the flow. The
second term, compared to numerical
results on Fig.\ref{pressureheq},  is related to the foam deformation.

In contrast to the deformation of the foam and with the resulting tension
forces, the pressure cannot
be measured locally from images of a 2D foam. As shown on Fig.
\ref{pressureheq}, it may nevertheless be very
heterogeneous and plays a crucial role in the foam dynamics. It is thus an
experimental challenge to
determine the pressure in each bubble in a flowing foam.

The same approach can in principle be used to measure the displacement field.
We thus compared the position of the bubble center before and after relaxation.
The small compressibility of the foam induces
a mean parabolic displacement of the form $X_x = \lambda (x^2-L^2)$ because the
bubble volumes increase upstream
and decrease  downstream during relaxation. Even with very small 
variations of bubble areas, as the displacement
is the integral of the deformation, this effect is larger than the 
small elastic displacement we want to
study. If this contribution is subtracted, the displacement field is 
nevertheless qualitatively different from
the purely elastic displacement field computed with the software 
Freefem, for the boundary conditions obtained
with the vertex model simulations.
More simulations are needed to conclude to the physical relevance of 
such results and to exclude that they are just artefacts.

\section{Summary and conclusion}
\label{summary}

In conclusion, we have shown that the {\it large bubble instability} 
can be reproduced with
simple numerical simulations, keeping only the skeleton of the foam 
structure and the three key
  ingredients which are tension forces,
viscous forces and volume conservation in the bubbles. The main 
physical aspects of this dynamical
behavior is understood as a complex interplay between the viscous, 
elastic and plastic properties of the
foam, seen as a continuum medium.
This theoretical approach leads to a good agreement with numerical 
and experimental results.
It can be adapted for various situations, as
  foam flows around obstacles for instance.  Future work will have to 
take into account the disordered (or  possibly
  crystallized) foam organization to deal with the velocity 
fluctuations of the large bubble.

  A very important perspective is to predict the flow properties of a 
fully polydisperse foam. A mean field theory
will probably be insufficient, as bubbles seem to be strongly coupled 
and as the flowing foam reorganizes itself with a resulting size 
segregation effect.

{\bf Acknowledgments}:
We thank {\it Rennes M\'etropole} and CNRS for financial support and 
G. Le Ca\"er, J. Lambert, S. Cox, F. Graner and B. Dollet  for 
enlightening discussions.

\appendix

\section{Hydrodynamical analogy}
\label{analogie_hydro}
Hydrodynamics equations for an incompressible viscous fluid of velocity $v_h$,
pressure $P_h$  and viscosity $\eta$ is :
\begin{equation}
\eta \Delta \vec{v}_h - \nabla P_h =0 \; ; \; \nabla \cdot \vec{v}_h =0 \; .
\label{eq_hydro}
\end{equation}
The problem of an infinite cylinder of diameter $D$ pushed at constant velocity
$V_h$ between two walls
separated by a distance $2L$ has been solved by Faxen \cite{faxen38}.
The variables can be identified as follows
  $ \eta {\bf v}_h \leftrightarrow \gamma h / d \; {\bf X}$, $P_h 
\leftrightarrow
\eta v_0 x
/d + \bar{P}$ and $V_h \leftrightarrow  \delta$.
  The expression of the force exerted on the cylinder is \cite{faxen38,happel}
\begin{eqnarray}
&& \int_{\cal C} - P_h n_x + \eta \left ( 2 \drp{v_{hx}}{x} n_x +
\left(\drp{v_{hy}}{x} +\drp{v_{hx}}{y}\right) n_y  \right) ds \nonumber \\
&=& -{4 \pi \eta V_h  \over ln({2 L \over D}) - 0.9157 + 1.73 \, {D^2 \over (2
L)^2} + ... }
\label{forcehydro}
\end{eqnarray}
Using the variable identification, we obtain
\begin{eqnarray}
&&\int_{\cal C} - ( \eta v_0 x /d + \bar{P}) n_x + \nonumber\\
&&\gamma h / d  \,  \left( 2 \drp{X_x}{x} n_x +
\left(\drp{X_y}{x} + \drp{X_x}{y}\right) n_y  \right) ds \nonumber\\
&=& -{4 \pi\, \gamma\, h \,\delta   \over d \,\left(ln({2 L \over D}) - 0.9 +
1.73 \, {D^2 \over (2 L)^2} + ...\right) }
\end{eqnarray}
The term we need to compute is given by eq. \ref{drivingforce} and is thus
\begin{equation}
F_x = \int_{\cal C}{ \eta v_0 x \over d } dx-{4 \pi {\gamma h \over d}  \delta
\over ln({2L \over D}) - 0.9 + 1.73 \, {D^2 \over (2 L)^2} +...}
\label{drivingforce2b}
\end{equation}

\section{Resolution of eq. \ref{eq_mouv}}

\label{sol_exacte}
Equation  \ref{eq_mouv} is adimensional with the time unit $\tau = D \, d \,
Ca_0^{2/3}/h v_0$.
With,  $\bar{t}= t/\tau$ and $\bar v = v/v_0$ and omitting  the bar
notation, we
get,

\begin{equation}
\int_0^t (v-1) dt = {D \over d} - {3 \over 2} v^{2/3}
\label{eq_mouv_adim}
\end{equation}
whose differentiation yields
\begin{equation}
v^{4/3}-v^{1/3}+ {dv \over dt}=0
\label{eq_mouv_adim_der}
\end{equation}
The solution of this differential  equation with separable variables is
\begin{equation}
t=H(v(t))-H(v(0))
\end{equation}
with
\begin{equation}
H(x) = \ln \left({\sqrt{x^{2/3}+x^{1/3}+1} \over x^{1/3}-1} \right)-\sqrt{3}
\tan^{-1}\left ( {2 x ^{1/3} +1 \over \sqrt{3}}  \right )
\end{equation}
The initial velocity is $v(0) = (D/d)^{3/2}$ and the maximum time
before a T1 is
given by
\begin{equation}
\int_0^{t_m} (v-1) dt ={D \over d} - {3 \over 2} v(t_m)^{2/3} = {D \over v_0
\tau}
\end{equation}

\begin{equation}
v(t_m) = {2 \over 3} \left ( {D \over d} - {D \over v_0 \tau} \right )^{3/2}
\end{equation}
We finally express the mean relative velocity of the large bubble in terms of
adimensional  variables as:
\begin{equation}
<v-1> ={D \over v_0 \tau} {1 \over H(v(t_m))-H((D/d)^{3/2}) }
\label{anal_exact}
\end{equation}

\end{document}